\documentclass[aps,twocolumn,showpacs,floatfix,amssymb,preprintnumbers,pre]{revtex4-1}

\usepackage{graphicx}
\usepackage{dcolumn}
\usepackage{bm}
\usepackage{epsfig}
\usepackage{amsmath}
\usepackage{amssymb}
\usepackage{amsfonts}
\usepackage{latexsym}
\usepackage{color}

\newcommand{\Ds}{D_{\rm s}}
\newcommand{\Rs}{R_{\rm s}}
\newcommand{\muI}{\mu(I)}
\newcommand{\Do}{D_{\rm o}}
\newcommand{\As}{A_{\rm s}}
\newcommand{\Ao}{A_{\rm o}}
\newcommand{\rhob}{\rho_{\rm b}}
\newcommand{\rhoo}{\rho_{\rm o}}
\newcommand{\Wd}{\dot{W}_{\rm d}}
\newcommand{\msr}{\langle\sigma_{rr}\rangle}
\newcommand{\szzo}{\sigma^{\rm bott}_{zz}}

\begin{document}

\title{Differential equation for the flow rate of discharging silos based on energy balance}

\author{J. R. Darias$^{1}$, Marcos A. Madrid$^{2,3}$, Luis A. Pugnaloni$^{4}$}
\affiliation{$^{1}$Laboratorio de \'Optica y Fluidos, Universidad Sim\'on Bol\'ivar, Apartado Postal 89000, Caracas 1080-A, Venezuela.\\
$^{2}$Departamento de Ingenier\'ia Mec\'anica, Facultad Regional La Plata, Universidad Tecnol\'ogica Nacional, CONICET, Av. 60 Esq. 124, 1900 La Plata, Argentina.\\$^{3}$Instituto de F\'isica de L\'iquidos y Sistemas Biol\'ogicos (CONICET La Plata, UNLP), Calle 59 Nro 789, 1900 La Plata, Argentina.\\$^{4}$Departamento de F\'isica, Facultad de Ciencias Exactas y Naturales, Universidad Nacional de La Pampa, CONICET, Uruguay 151, 6300 Santa Rosa (La Pampa), Argentina.}

\begin{abstract}
Since the early work of Hagen in 1852 and Beverloo et al. in 1961, the flow rate of granular material discharging through a circular orifice from a silo has been described by means of dimensional analysis and experimental fits, and explained through the ``free fall arch'' model. Here, in contrast with the traditional approach, we derive a differential equation based on the energy balance of the system. This equation is consistent with the well known Beverloo rule thanks to a compensation of energy terms. Moreover, this new equation can be used to explore new conditions for silo discharges. In particular, we show how the effect of friction on the flow rate can be predicted. The theory is validated using discrete element method simulations.
\end{abstract}
\maketitle

\section{Introduction}
\label{intro}

When a granular material (such as seeds) flow through an orifice at the base of a silo, the resulting flow rate presents peculiar features when compared with the better known phenomenology of inviscid fluids. In particular, the flow rate of grains does not depend on the height $h$ of the material in the silo. This was early noted by Hagen in 1852 \cite{hagen1852} (for a translation of the German publication see Ref. \cite{tighe2007}). Moreover, Hagen showed that the mass flow rate $Q$ scales as $D_\textrm{o}^{5/2}$, with $D_\textrm{o}$ the diameter of the circular orifice. Compare this with $D_\textrm{o}^2$, which is valid for inviscid fluids. Hagen also provided the first heuristic postulate to explain this 5/2 power which was later termed the ``free fall arch'' model \cite{BrownBook}. In brief, Hagen postulated that grains move downwards in the silo at a very low velocity until they arrive at a region (about one orifice radius tall) close to the opening at which the local density is low and grains perform a simple free fall. From here, the typical vertical velocity $v$ at which grains cross the plane of the orifice can be estimated ($v_\textrm{out}=\sqrt{g D_\textrm{o}}$, with $g$ the acceleration of gravity) and the mas flow rate calculated as $Q = \pi (D_\textrm{o}/2)^2 \rho_\textrm{o} v_\textrm{out} = (\pi/4) \rho_\textrm{o} \sqrt{g} D_\textrm{o}^{5/2}$, being $\rho_\textrm{o}$ the local apparent density at the orifice. It is customary to replace $\rho_\textrm{o}$ by the density in the bulk of the silo $\rho_\textrm{b}$. Therefore, the $\pi/4$ coefficient is replaced by a constant $C$ that is later fitted to the experimental data. Hagen also noted that this expression does not agree entirely well with the data. The solution to this was to consider that the effective orifice is about one grain diameter, $d$, smaller due to boundary effects which leads to $Q = C \rho_b \sqrt{g} (D_\textrm{o}-d)^{5/2}$.

In 1961 Beverloo et al. carried out a series of experiments and found a more suitable expression to fit the mass flow rate \cite{beverloo}. This expression (see below) became widely accepted, and is usually named as the Beverloo rule. Beverloo et al. did not refer to Hagen's work, but found the 5/2 power based on dimensional analysis. Others in the decade preceding Beverloo proposed expression that were less successful. The improvement with respect to Hagen was the introduction of an additional constant $k$ to the boundary effect correction. The Beverloo rule states that the mass flow rate is   

\begin{equation}
Q = C \rho_\textrm{b} \sqrt{g}(D_\textrm{o}-k\,d)^{5/2}, \label{beverloo}
\end{equation}
where $k$ and $C$ are two fitting dimensionless constants. The fitted value for $k_\textrm{experim}$ may vary between $1.4$ and $3$ depending on the shape and size dispersion of the grains. However, $C_\textrm{experim}\approx 0.58$ with almost no influence of the type of material the grains are made of \cite{NeddermanBook}. The origin for this ``universal'' value of $C$ did not receive much attention until recently \cite{madrid2018}.

The problem of discharge of grains through an orifice has been revisited by a number of authors (see for example \cite{nedderman1982,tuzun,savage,ristow,mankoc2007,kondic2014,wilson2014,goldberg2016,villagran} and references therein). The basic phenomenology has been confirmed in all studies, i.e.: (i) the flow rate does not depend on the column height, (ii) the flow rate scales with $D_\textrm{o}^{5/2}$, (iii) the prefactor is $C\approx 0.58$ for almost all materials. However, Rubio-Largo et al. provided evidence that the concept of ``free fall arch'' may not be a realistic picture of the internal dynamics in the silo around the orifice \cite{rubio2015}.

This mechanical problem of discharge should be described from first principles via energy balance as it is done for a fluid (e.g., Bernoulli's law). To our knowledge, this has not been done successfully so far, despite some attempts (see for example \cite{mcdougall1965}). One possible reason for this is that the rheological response of the grains while flowing in a silo has received less attention, leaving a gap in a basic component of the energy balance analysis: \emph{the dissipated power}. In this direction, Staron et al. \cite{staron2012} have proved that the flow in a discharging silo can be described by introducing the $\muI$-rheology model \cite{dacruz2005} into the Navier--Stokes equations. This is, in practice, the introduction of an effective dissipation term into the equations. Recently, we have shown that, at a global scale in the silo, the flow is consistent with the quasistatic limit in the $\muI$-rheology \cite{madrid2018}. In that work, we have provided an expression for the energy dissipated during the discharge of a silo that will be revisited in what follows.

In this work, we derive a differential equation for $Q$ as a function of time during the discharge. This is achieved by using the work--energy theorem for the system of interest, which is defined as the set of grains that remain inside the silo at any given time. To calculate the energy dissipation we use a revised expression to the one presented in Ref. \cite{madrid2018}. The result is consistent with the  Beverloo rule. We provide a theoretical estimate for the value of $C$ that is remarkably close to the value obtained by experimental fits. Moreover, the new equation predicts an increase in the flow rate when the friction coefficient is reduced, which is consistent with simulation results. 

\section{System definitions}

\begin{figure}
 \begin{center}
  \includegraphics[width=0.6\columnwidth]{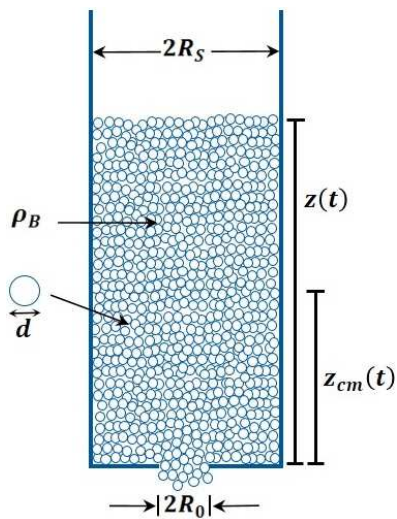}
 \end{center}
\caption{Sketch of the axial cross section of a cylindrical silo.}
\label{set-up}
\end{figure}

We consider a cylindrical silo (see Fig. \ref{set-up}) of diameter $\Ds$ (radius $R_\textrm{s}=\Ds/2$ and cross section $A_\textrm{s}=\pi R_\textrm{s}^2$) with a flat base. This base has an orifice of diameter $D_\textrm{o}$ in its center (radius $R_\textrm{o}=D_\textrm{o}/2$ and cross section $A_\textrm{o}=\pi R_\textrm{o}^2$). With the orifice covered, the silo is filled with an initial mass $M_\textrm{ini}$ of a granular material that fills the silo up to a certain height. We will consider that the bulk apparent density of the material $\rho_\textrm{b}$ is homogeneous throughout the column as a first order approximation. The mass of each grain is denoted by $m$. 

When the orifice is opened, the discharge of grains starts. During the discharge, the mass $M(t)$ inside the silo at time $t$ can be written as
\begin{equation}{\label{equiv}}
 M(t)=\rho_\textrm{b} A_\textrm{s} z(t)=2\rho_\textrm{b} A_\textrm{s} z_\textrm{cm}(t),
\end{equation}
where $z(t)$ is the height of the column of grains, and $z_\textrm{cm}(t)=z(t)/2$ is the center of mass of the granular column at time $t$. Therefore, the mass flow rate $Q(t)$ is
\begin{equation}
 Q(t)=-\dot{M}(t)=2 \rho_\textrm{b} A_\textrm{s} v_\textrm{cm}(t), \label{q-vcm}
\end{equation}
where $v_\textrm{cm}(t)=|\dot{z}_\textrm{cm}(t)|$ is the speed of the center of mass of the granular column. Note that since $M(t)$ decreases with time during the discharge, $\dot{M}$ and $\dot{z}_\textrm{cm}$ are negative.

\section{Simulations}

To validate our theoretical model we carried out a series of DEM simulations of spherical particles in a cylindrical silo as described in the previous section. We used the LIGGGHTS \cite{liggghts} implementation with a particle--particle Hertz interaction and Coulomb criterion using a Young modulus $Y=70$ MPa, Poison ratio $\nu=0.25$, restitution coefficient $e=0.95$ and friction coefficient $0.1 <\mu < 1.0$. The same interaction applies for the particle--walls contacts. Details on the particle--particle interactions are given in Appendix A. The particle diameters are $d=1$ mm and their material density is $\rho = 2500\ \rm{kg/m}^3$. The silo diameter is $D_\mathrm{s} = 30\ \rm{mm}$ (some test have been run also with $D_\mathrm{s} = 24\ \rm{mm}$). The orifice diameter is varied in the range $6.0\ {\rm mm} \leq\Do\leq 10.0\ {\rm mm}$. Particles are poured into the silo to fill a height $z(t=0)\approx 10 D_\mathrm{s}$ (which implies up to $3 \times 10^5$ grains, depending on the silo diameter). The orifice is initially blocked by a plug. We let the grains come to rest in the silo by waiting until the kinetic energy per particle falls below $10^{-10}$ J. Then, we remove the plug and allow the material to discharge through the orifice. {Although particles are monosized, we do not observe crystalline structures in the simulations.} The magnitude of the acceleration of gravity is set to $g=9.81$ m/s$^{2}$ and the integration time step is $\Delta t=5\times10^{-6}$ s.

\section{Work--energy theorem and dissipated power}
\label{model}

We focus on the system composed of the grains inside the silo at any given time. According to the work--energy theorem, the change in kinetic energy $\dot{K}_\textrm{in}$ of this system of grains is

\begin{equation}
\dot{K}_\textrm{in} = \dot{W}_\textrm{g} - \dot{W}_\textrm{out} + \dot{W}_\textrm{el} - \dot{W}_\textrm{d}, \label{energy-balance}
\end{equation}
where $\dot{W}_\textrm{g}$ is the power injected by the force of gravity acting on the grains, $\dot{W}_{out}$ is the power loss due to the grains that leave the silo through the orifice at a velocity $v_\textrm{out}$, $\dot{W}_\textrm{el}$ is the ``elastic power'', i.e.,  the rate of change of the elastic energy of the grain--grain contacts, and $\dot{W}_\textrm{d}$ is the dissipated power due to the non-conservative interactions (friction and inelastic collisions between the grains and between the grains and the walls). Some of the terms in Eq. (\ref{energy-balance}) are in fact negligible ($\dot{K}_\textrm{in}$ and $\dot{W}_\textrm{el}$) and some are easy to calculate from basic mechanics ($\dot{W}_\textrm{g}$ and $\dot{W}_\textrm{out}$). We discuss those contributions in Appendix B. In the remaining of this subsection we focus on the dissipated power $\dot{W}_\textrm{d}$.

In a recent work \cite{madrid2018}, we have shown that the power dissipated during a silo discharge can be calculated by assuming that the flow is consistent with a quasistatic shear flow in the framework of the $\muI$-rheology \cite{dacruz2005}. Here, we present an improved expression for $\dot{W}_\textrm{d}$ in which we consider in more detail the local dissipation due to the convergent flow region in the bottom part of the silo.

\begin{figure}
 \begin{center}
  \includegraphics[width=0.6\columnwidth]{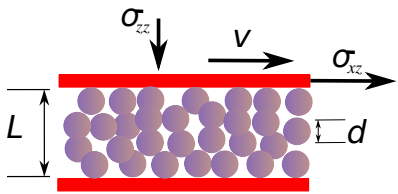}
 \end{center}
\caption{Sketch of a simple plane shear cell. The top plate is driven at constant velocity $v$.}
\label{simple-shear}
\end{figure}

Let us first consider a simple plane shear configuration (see Fig. \ref{simple-shear}). 
According to the $\muI$-rheology model \cite{dacruz2005}, the average tangential stress $\sigma_{xz}$ required to keep the top plate moving at constant velocity $v$ can be written as $\sigma_{xz}= \muI \sigma_{zz}$. Here, $\sigma_{zz}$ is the confining stress and $\muI$ is an ``effective friction coefficient'' that accounts for all the complex interactions in the granular sample between the plates. The effective friction depends on the inertial number $I=v d \sqrt{\rho}/ (L \sqrt{\sigma_{zz}})$, with $\rho$ the density of the material the grains are made of \cite{dacruz2005}. The inertial number suffices to characterize the flow as long as the grains are stiff and $L \gg d$. 

The power $\dot{W}_{\rm d}$ dissipated during the motion of the top plate in Fig. \ref{simple-shear} is simply given by $\dot{W}_{\rm d}=\sigma_{xz} A v$, where $A$ is the total area of the top plate. Hence, the dissipated power can be written as $\dot{W}_{\rm d}=\muI \sigma_{zz} A v$. In this expression, the effect of the properties of the granular material on the dissipation comes only through $\muI$ since $\sigma_{zz}$, $A$ and $v$ are control variables of the experiment. Interestingly, it has been shown that $\muI$ follows a universal curve for all granular materials as long as the particle--particle friction is high enough (roughly above $0.4$) \cite{dacruz2005}. However, this curve depends on geometrical factors like dimensionality \cite{azema2014}. In the quasistatic limit (i.e., $I\ll1$, with $I<0.01$ being a typical criterion used for practical proposes), the value of $\mu(I=0)$ becomes unique to all granular materials for a given geometry if the particle--particle friction is above $0.4$. Therefore, the dissipated power becomes independent from the details of the particle--particle interactions in this limit (i.e., when $I \rightarrow 0$ and $\mu > 0.4$).     

We postulate that the flow inside a discharging silo can be modeled in a similar way as in the simple plane shear geometry. In the cylindrical silo geometry, up and away from the converging flow observed at the bottom of the silo, the confining pressure can be taken as the mean radial pressure $\msr(t)$ (where the angular brackets indicate an average over the height of the granular column), the driving velocity as the velocity of the free surface of the granular column $v=2v_\textrm{cm}$ and the ``plate area'' as the area of contact between the grains and the silo lateral walls [$A_1(t)=\pi \Ds 2 z_\textrm{cm}(t)$]. {Note that $v$ is only a velocity that is characteristic for the motion of the system at the macroscopic scale (as the plate velocity in the plane shear experiment) and does not need to be compared with the actual velocities of the grains nor with the velocity gradients in the system.}

Close to the bottom of the silo, over a height of about $\Rs$, the flow converges to the orifice and the relevant confining pressure is no longer radial in this section of the silo. We therefore use, as a proxy for the confining pressure in this region, the vertical component $\szzo(t)$ averaged over the bottom part of the silo (from the base to a height $\Rs$). The lateral area of this region is $A=\pi \Ds \Rs$, and we use also the characteristic velocity $v=2v_\textrm{cm}$. However, there exist a small region of height $\approx R_\textrm{o}$ right above the base that does not contribute to the dissipation of energy. This is due to the fact that grains close to the solid base do not move significantly and then do not dissipate energy. Of course, grains in the neighborhood of the orifice do move at high velocities, but here the packing fraction is so low that there are very few particle--particle interactions to dissipate energy. Therefore, for the converging flow zone we take the effective lateral area as $A_2= 2 \pi \Rs (\Rs - \Do/2)$. Using these approximations we can write the dissipated power in the entire granular column as

\begin{equation}
 \dot{W}_\textrm{d}(t)= \muI [ \msr(t) A_1(t)  + \alpha \szzo(t) A_2 ] 2 v_\textrm{cm}(t). \label{wd1}
\end{equation}
The first term in Eq. (\ref{wd1}) accounts for the dissipation along the upper part of the silo, whereas the second term accounts for the contribution of the converging flow. The constant $\alpha$ in this second term is introduced as a correction factor since the active area, characteristic velocity and confining pressure in the converging zone are only estimates that should give the correct scaling on the various variables but not necessarily the correct numerical value for the dissipation. 

Using Eqs. (\ref{equiv}) and (\ref{q-vcm}), we can write Eq. (\ref{wd1}) as

\begin{align}
\dot{W}_\textrm{d}(t) = &- \muI 2 \pi \Rs \left[ \msr(t)  \frac{M(t)}{\rhob \As} \right. \notag \\
&+ \left. \alpha \szzo(t) \left(\Rs-\frac{\Do}{2}\right) \right] \frac{\dot{M}(t)}{\rhob \As}. \label{wd}
\end{align}

\section{A differential equation for the mass flow rate}\label{sec-diff}

Collecting all terms for the energy balance from Appendix B and the previous section, i.e., plugging Eqs. (\ref{k-in}), (\ref{wg}), (\ref{wout}), (\ref{elastic}) and (\ref{wd}) into Eq. (\ref{energy-balance}), one can obtain a differential equation for the mass flow rate $Q\equiv-\dot{M}(t)$. As we mentioned in Appendix B, some of these terms are in fact negligible. In particular, we take $\dot{K}_\textrm{in}=0$ and $\dot{E}_\textrm{el}=0$. Finally, Eq. (\ref{energy-balance}) can be written as 

\begin{align} \label{pre-diff}
 0 =  &-\frac{g}{\rhob \As} M(t) \dot{M}(t) 
       + \frac{\dot{M}^3(t)}{2\rhoo^2 \Ao^2} + \muI 2 \pi \Rs\\
   &\times  \left[ \msr(t) \frac{M(t)}{\rhob \As}+ \alpha \szzo(t) \left(\Rs-\frac{\Do}{2}\right) \right] \frac{\dot{M}(t)} {\rhob \As}. \notag
\end{align}

Solving for $-\dot{M}$ \cite{foot}

\begin{eqnarray}  \label{diff}
-\dot{M}(t) =&& \frac{\pi\sqrt{2}}{4} \rhoo \sqrt{g} \Do^2 \left\lbrace \frac{M(t)}{\rhob \As} - \frac{\muI 2 \pi \Rs}{g \rhob \As} \right. \\
  &\times& \left. \left[\msr(t) \frac{M(t)}{\rho_{\rm b} A_{\rm s}}+ \alpha \szzo(t)  (\Rs-\frac{\Do}{2})\right] \right\rbrace ^{1/2}.\notag
\end{eqnarray}

Equation (\ref{diff}) is a first order differential equation for $M(t)$ that can be closed with an initial condition such as $M(t=0)= M_0$. To solve this equation it is necessary to know $\msr(t)$, $\szzo(t)$, $\muI$ and $\alpha$. In the next section we revise the Walters model for the pressure in discharging silos to obtain analytical expressions for $\msr(t)$ and $\szzo(t)$ to close Eq. (\ref{diff}).

It is worth mentioning at this point that Eq. (\ref{pre-diff}) reduces to the equation for an inviscid fluid if the last term that accounts for the dissipated power is neglected. In this case we obtain

\begin{eqnarray}  \label{bernoulli}
-\frac{\dot{M}^2(t)}{\rho^2_\textrm{o} A^2_\textrm{o}} &=& 2 g \frac{M(t)}{\rho_\textrm{b} A_\textrm{s}},\notag\\
v_{\rm out}&=& \sqrt{2 g z(t)},
\end{eqnarray}
where $z(t)$ is the column height and $v_{\rm out}=\dot{M}(t)/(\rho^2_\textrm{o} A^2_\textrm{o})$ is the mean velocity of the outflowing material.

\section{Pressure in a discharging silo}

\subsection{Walters dynamic stresses}

Walters \cite{walters1973} and Walker \cite{walker1966} have developed models for the pressure during silo discharge following an approach similar to Janssen's. In order to close Eq. (\ref{diff}) we will use these previous developments. The expression for $\sigma_{zz}$ and $\sigma_{rr}$ at a given depth $z'$ (measured from the free surface of the granular column) into the moving column is

\begin{align}
 \sigma_{zz}^{\rm Walters}(z') &=  \frac{g \rhob \Ds}{4 B} [1 - {\rm e}^{-4 B z'/\Ds}], \\
 \sigma_{rr}^{\rm Walters}(z') &=  \frac{g \rhob \Ds}{4 \tan\phi} [1 - {\rm e}^{-4 B z'/\Ds}],\label{eq-walters}
\end{align}
where $\tan\phi$ is the effective friction corresponding to the wall yield locus \cite{NeddermanBook}, and $B$ plays the role of the well-known Janssen's force redirection factor. For discharging (not static) silos Walters obtains \cite{walters1973}

\begin{align}\label{eq-B}
 B &= \frac{\tan\phi \cos^2\delta}{(1+\sin^2\delta)-2y\sin\delta}, \\
 y &= \frac{2}{3c}[1-(1-c)^{3/2}],\notag\\
 c &= \frac{\tan^2\phi}{\tan^2\delta}.\notag 
\end{align}
Here, $\phi$ is the effective friction angle for the wall yield locus and $\delta$ is the effective friction angle for the internal yield locus. As discussed by Nedderman (see section 3.7 in Ref. \cite{NeddermanBook}), the wall yield locus $\phi$ should not be set simply as ${\rm atan}(\mu_{\rm wall})$. Instead, the Jenike's rule should be applied

\begin{equation}
 \tan\phi = \left\lbrace {\begin{array}{l} \mu_{\rm wall} \text{ if } \sin\delta \geq \mu_{\rm wall} \\ \sin\delta \text{ if } \sin\delta < \mu_{\rm wall} \end{array}}  \right. . \label{ec-jenike}
\end{equation}

\begin{figure}
 \begin{center}
  \includegraphics[width=0.8\columnwidth]{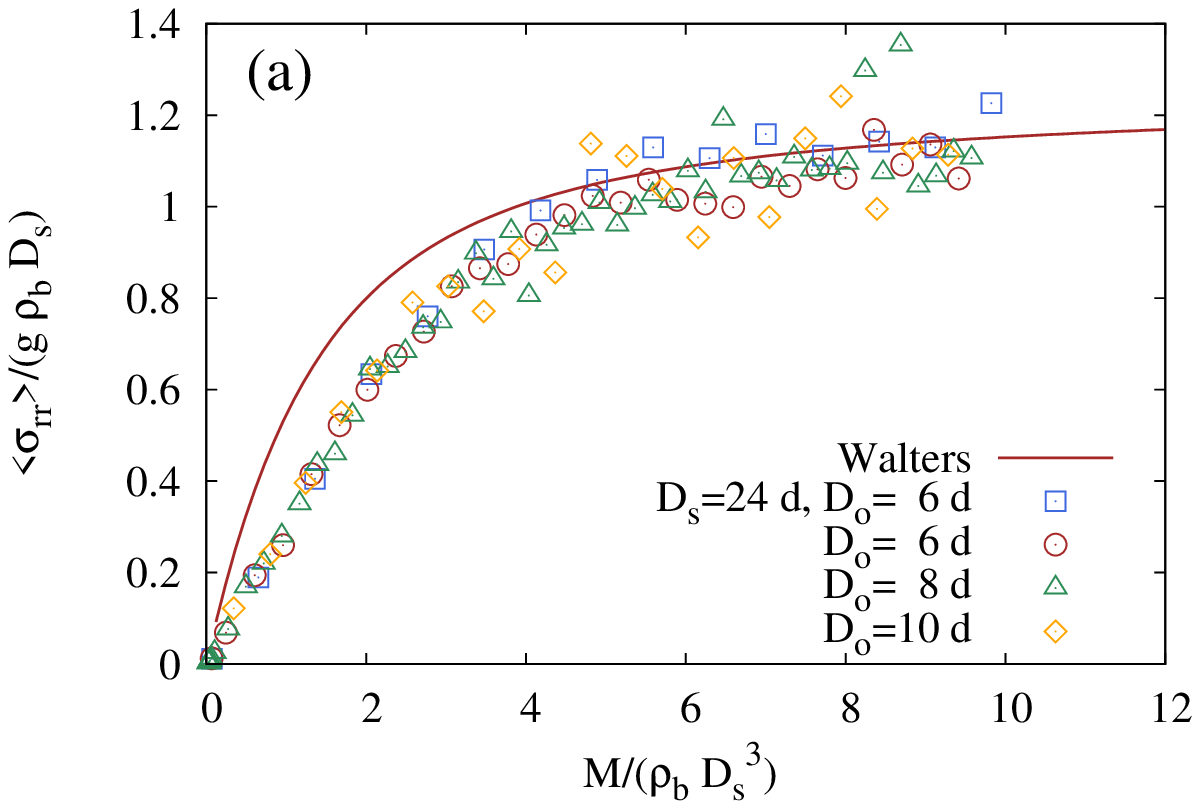}\\
  \includegraphics[width=0.8\columnwidth]{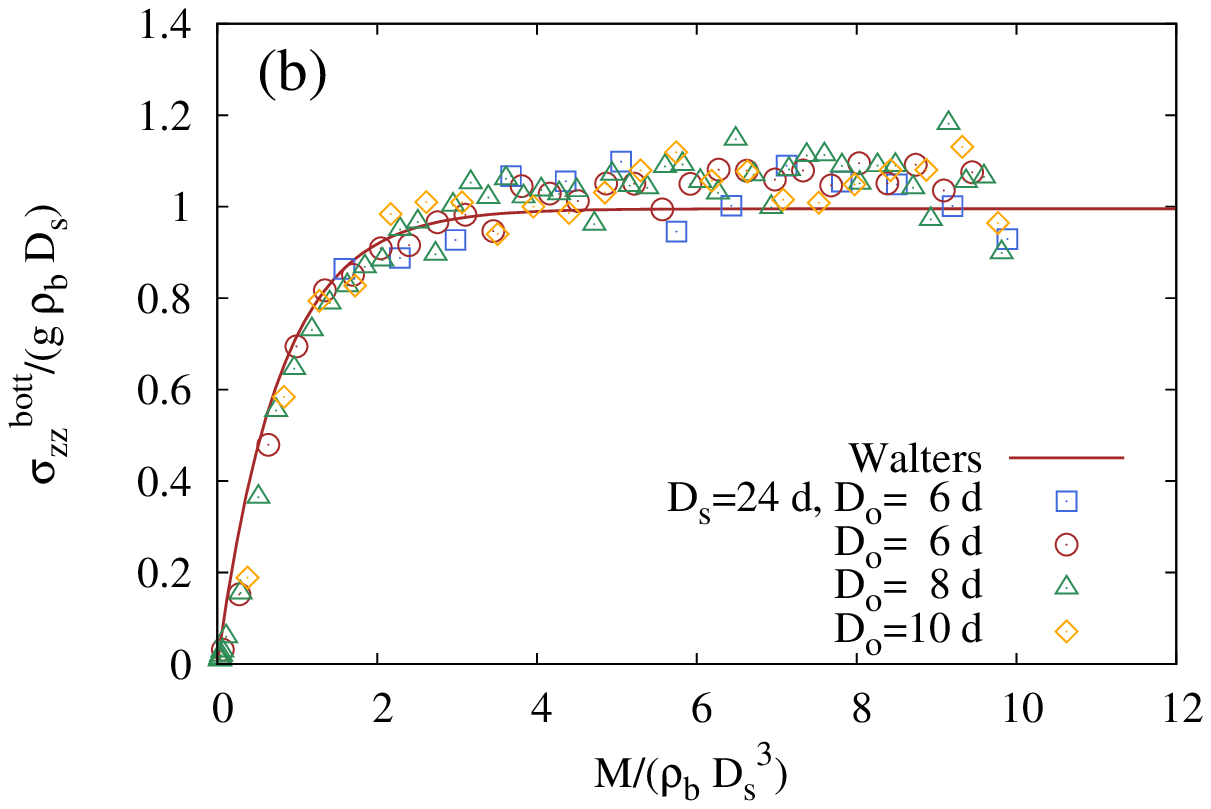}
 \end{center}
\caption{(a) $\msr$ as a function of the total mass $M$ in the silo. (b) $\szzo$ as a function of the total mass $M$ in the silo. Symbols correspond to DEM simulations for different orifice diameters $\Do$ and different silo diameters $\Ds$. Unless otherwise stated $\Ds=30d$. The particle--particle and particle--wall friction coefficients were set to $\mu = 0.5$. The solid lines correspond to: (a) $\msr$ according to Eq. (\ref{eq-walters2}) and (b) $\szzo(z)$ given by Eq. (\ref{eq-szzo}). We have set $\tan\delta=0.204$ to fit simultaneously both pressures. For $\tan\phi$ we used $\tan\phi=\sin\delta$, following Eq. (\ref{ec-jenike}).}
\label{fig-presion}
\end{figure}

For the vertical pressure on the converging flow zone we will use as an estimate

\begin{align} \label{eq-szzo}
\szzo(z) &= \sigma_{zz}^{\rm Walters}(z'=z)\\ \notag
&=\frac{g \rhob \Ds}{4 B} [1 - {\rm e}^{-4 B z/\Ds}],
\end{align}
being $z$ the granular column height. Note that Walters defines $z'$ in the negative direction, therefore  $\sigma_{zz}^{\rm Walters}(z'=z)$ is simply the vertical pressure on the base of the silo.

Since in Eq. (\ref{diff}) we also require the average $\msr$ over the entire column of grains, we average Eq. (\ref{eq-walters}) from $z'=0$ to the total depth $z'=z$ of the column, which yields

\begin{align}\label{eq-walters2}
\msr(z)&= \frac{1}{z} \int_{0}^{z}{\sigma_{rr}^{\rm Walters}(z')dz'} \notag\\
&= \frac{g \rhob \Ds}{4 \tan\phi} \left[1+ \frac{{\rm e}^{-4 B z/\Ds}-1}{4 B z/\Ds} \right]. 
\end{align}

In Fig. \ref{fig-presion} we show $\szzo$ and $\msr$  obtained from DEM simulations along with the theoretical predictions form Eq. (\ref{eq-szzo}) and (\ref{eq-walters2}), respectively. The particle--wall friction was set $\mu_{\rm wall}=0.5$, which is larger than the internal friction of the material. The curves have been fitted setting $\tan\delta=0.204$ and $\tan\phi=\sin\delta$, following Eq. (\ref{ec-jenike}). 

There are two interesting features to emphasize in Fig. \ref{fig-presion}. Firstly, $\msr$ saturates much more slowly than the bottom pressure $\szzo$. Secondly, the prediction for $\msr$ fails to some extent for low column heights. Although this could be improved, we will show that these expression for the pressure are sufficient to obtain new valuable insights into the silo discharge.

\subsection{Asymptotic pressure}

A first order approximation to the solution for Eq. (\ref{diff}) can be obtained by replacing the asymptotic limit of Eqs. (\ref{eq-szzo}) and (\ref{eq-walters2}) for high columns ($z \gg \Ds$). These asymptotic expressions are

\begin{align} \label{eq-asint}
  \szzo(z) &= \frac{g \rhob  \Ds}{4B} [1+O(z^2)]\\
  \msr(z) &= \frac{g \rhob  \Ds}{4\tan\phi} \left[ 1 - \frac{\Ds}{4 B z} + O(z^2)\right] \notag
\end{align}

\section{Comparison with the Beverloo rule}

\begin{figure}
 \begin{center}
  \includegraphics[width=0.8\columnwidth]{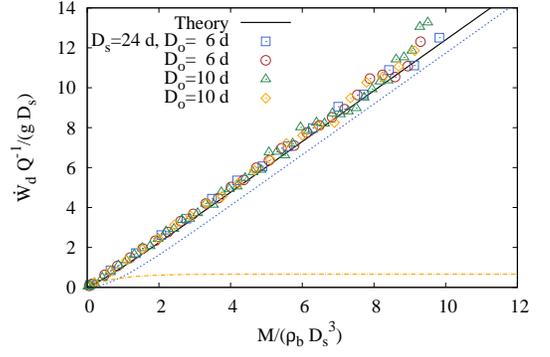}
 \end{center}
\caption{Dissipated power $\Wd$ (scaled by the mass flow rate $Q$) as a function of the total mass $M$ in the silo. Symbols correspond to DEM simulations for different orifice diameters $\Do$ and different silo diameters $\Ds$. If not stated $\Ds=30d$. The black solid line corresponds to the expression proposed in Eq. (\ref{wd}). The blue dashed and orange dot-dashed lines correspond, respectively, to the first and second term in Eq. (\ref{wd}) using $\muI=\tan\phi=\sin\delta=0.2$ and $\alpha=2.5$ as discussed in the text.}
\label{fig-disipacion}
\end{figure}

\subsection{The asymptotic equation and the $5/2$ power law}
Replacing Eq. (\ref{eq-asint}) into Eq. (\ref{diff}), and using $z = M(t)/(\rhob\As)$ (see Eq. (\ref{equiv})) we obtain

\begin{align} \label{eq-q-asint}
 -\dot{M}(t) =& \frac{\pi\sqrt{2}}{4} \rhoo \sqrt{g} \Do^2 \left[ \frac{M(t)}{\rhob \As} \left(1-\frac{\muI}{\tan\phi}\right) \right.\\
 &+ \left. \frac{\muI \Rs}{2B} \left(\frac{1}{\tan\phi}-2\alpha\right) + \frac{\alpha\muI}{2B} \Do \right]^{1/2} \notag
\end{align}

Equation (\ref{eq-q-asint}) becomes similar to the Hagen's expression for the flow rate if we select $\muI =  \tan\phi = (2\alpha)^{-1}$ and $B=1/4$. Under these conditions

\begin{equation} \label{eq-q-asint2}
 -\dot{M}(t) = \frac{\pi\sqrt{2}}{4} \rhoo \sqrt{g} \Do^{2.5}.
\end{equation}

As we showed in the previous section, the Walters expressions for the pressure fit the DEM data if $\tan\delta=0.204$ and $\tan\phi=\sin\delta=0.20$. For these values, Eq. (\ref{eq-B}) yields  $B=0.26 \approx 1/4$. Therefore, the well known Hagen's relation will hold if we simply set $\muI=0.20$ and $\alpha=2.5$.

To validate the values of $\muI$ and $\alpha$ required for Eq. (\ref{eq-q-asint}) to reduce to Eq. (\ref{eq-q-asint2}), we have calculated from our DEM simulations the energy dissipated during discharge {(see Appendix A for details)}. In Fig. \ref{fig-disipacion} we plot the dissipated power along with the expression provided by Eq. (\ref{wd}) using $\muI=0.20$ and $\alpha=2.5$. As we can see, the agreement is remarkable not only for high columns (where the asymptotic limit was used to predict $\muI$ and $\alpha$) but also for low column heights. This indicates that the two parameters introduced to model the dissipated power in practice take values that lead to a flow rate compatible with the Hagen's equation. It is worth mentioning that the contribution due to the second term in Eq. (\ref{wd1}), which corresponds to the dissipation in the converging zone of the flow, is small but not negligible (see dot-dashed line in Fig. \ref{fig-disipacion}).

\subsection{The bulk density and the prefactor $C$}

As we mentioned in the introduction, Hagen's prediction for the flow rate is $Q = -\dot{M} = (\pi/4) \rho_\textrm{o} \sqrt{g} D_\textrm{o}^{5/2}$. Therefore, by measuring the packing fraction $\rho_\textrm{o}$ at the opening and $\rho_\textrm{b}$ in the bulk of the silo one can provide an estimate for the constant $C$ since, according to Hagen's analysis, $\pi \rho_\textrm{o}/4 = C \rho_\textrm{b}$. During discharge, $\rho_\textrm{o} \approx \rho_\textrm{b}/2$ {(see Appendix C for details)}, implying that $C \approx 0.39$. Surprisingly, this estimate is more than $30 \%$ below the value obtained by fitting experiments by various author ($C_\textrm{experim}\approx 0.58$) \cite{NeddermanBook}. 

In Eq. (\ref{diff}) the non-dimensional prefactor (i.e., $\pi \sqrt{2}/4$) differs from the Hagen's prefactor.  If we use $\rho_\textrm{o}\approx \rho_\textrm{b}/2$ {(see Appendix C)} to replace the packing fraction at the orifice by the bulk packing fraction in Eq. (\ref{diff}), the non-dimensional prefactor becomes $\frac{\pi\sqrt{2}}{8} \approx 0.56$, which can be compared with the constant $C$ in the Beverloo rule. This value deviates less than $4 \%$ from the known value $C_\textrm{experim}\approx 0.58$. Hence our Eq. (\ref{diff}) seems to be much more accurate in predicting the prefactor that the Hagen equation.

{It is important to note that the value of $C$ is determined by the expression for the power lost through the orifice [see Eq. (\ref{wout}) and compare with Eqs. (\ref{pre-diff}) and (\ref{diff})]. A key role is played by the area of the cross section of the orifice $\Ao$. For a non-circular cross section, the factor $C$ will take a different value. For example, for a square orifice of side $L$ ($\Ao = L^2$), the prefactor becomes $C = \sqrt{2}/2$ (bear in mind that we take $\rho_\textrm{o} \approx \rho_\textrm{b}/2$). This is consistent with the experimental results from Ref. \cite{fowler1959}.}

Based in the previous discussion, in the rest of the paper we will set $C = \frac{\pi\sqrt{2}}{8}$ when we use the Beverloo expression for a circular orifice.

\subsection{The $-kd$ boundary effect correction}

When calculating our energy contributions in Appendix B, we assumed that the system can be considered as a continuum while calculating $\dot{W}_\textrm{out}$. We calculated the mass flow rate as $-\dot{M}(t)= \rho_\textrm{o} A_\textrm{o} v_\textrm{out}$ without considering that the orifice has a size only a few times the size of one grain. This introduces boundary corrections that we did not take into account.

\begin{figure}
 \begin{center}
  \includegraphics[width=0.8\columnwidth]{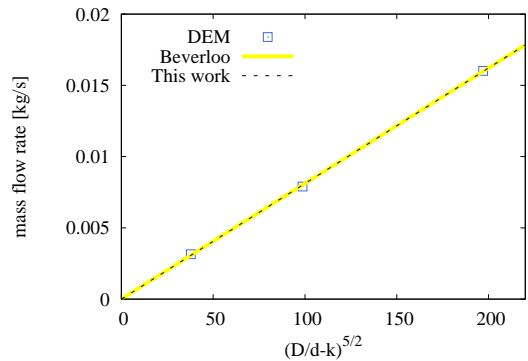}
 \end{center}
\caption{Mass flow rate as a function of $(D/d-k)^{5/2}$. Symbols correspond to the DEM data. The lines correspond to both the Beverloo equation {(yellow solid)} and the asymptotic expression {(black dashed)} for the present work (Eq. (\ref{eq-q-asint2}) plus the boundary effect correction) setting $C=\frac{\pi\sqrt{2}}{8}$ and fitting $k=1.72 \pm 0.02$. For these simulations $\Ds = 30d$ and $\mu=0.5$ for the particle--particle and the particle--wall interactions.}
\label{fig-D}
\end{figure}

The simplistic correction, in the style of Beverloo, done by replacing the orifice diameter $D_\textrm{o}$ by an effective smaller diameter $D_\textrm{o}-kd$ has been questioned \cite{mankoc2007,janda2012}. However, this is a simple way to incorporate this effect and we will use this in what follows. For spherical grains, most authors conducting either experiments or simulations in 3D indicate that $k \approx 1.4$. However, such fitting value corresponds to the parameter $C$ fitted to $0.58$. As discussed in the previous section, we now have a theoretical basis for setting $C=\frac{\pi\sqrt{2}}{8}$. If we do this and fit the only remaining parameter $k$ we find, for our DEM data (see Fig. \ref{fig-D}) that $k=1.72 \pm 0.02$. This is the value we will use for the rest of the paper. Notice that with $C$ and $k$ selected in this manner the Beverloo equation and the asymptotic expression (\ref{eq-q-asint2}) for the current theory coincide.

\section{Evolution of the flow rate during discharge}

\begin{figure}
 \begin{center}
  \includegraphics[width=0.8\columnwidth]{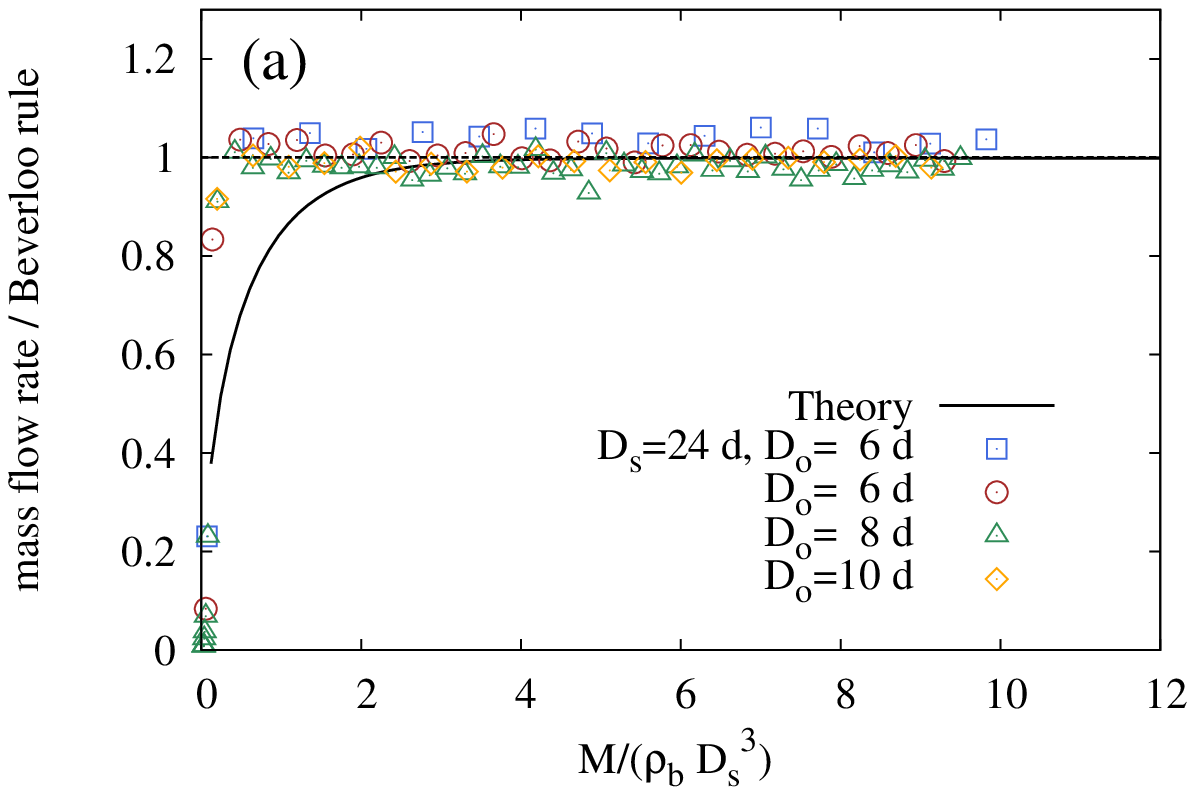}\\
  \includegraphics[width=0.8\columnwidth]{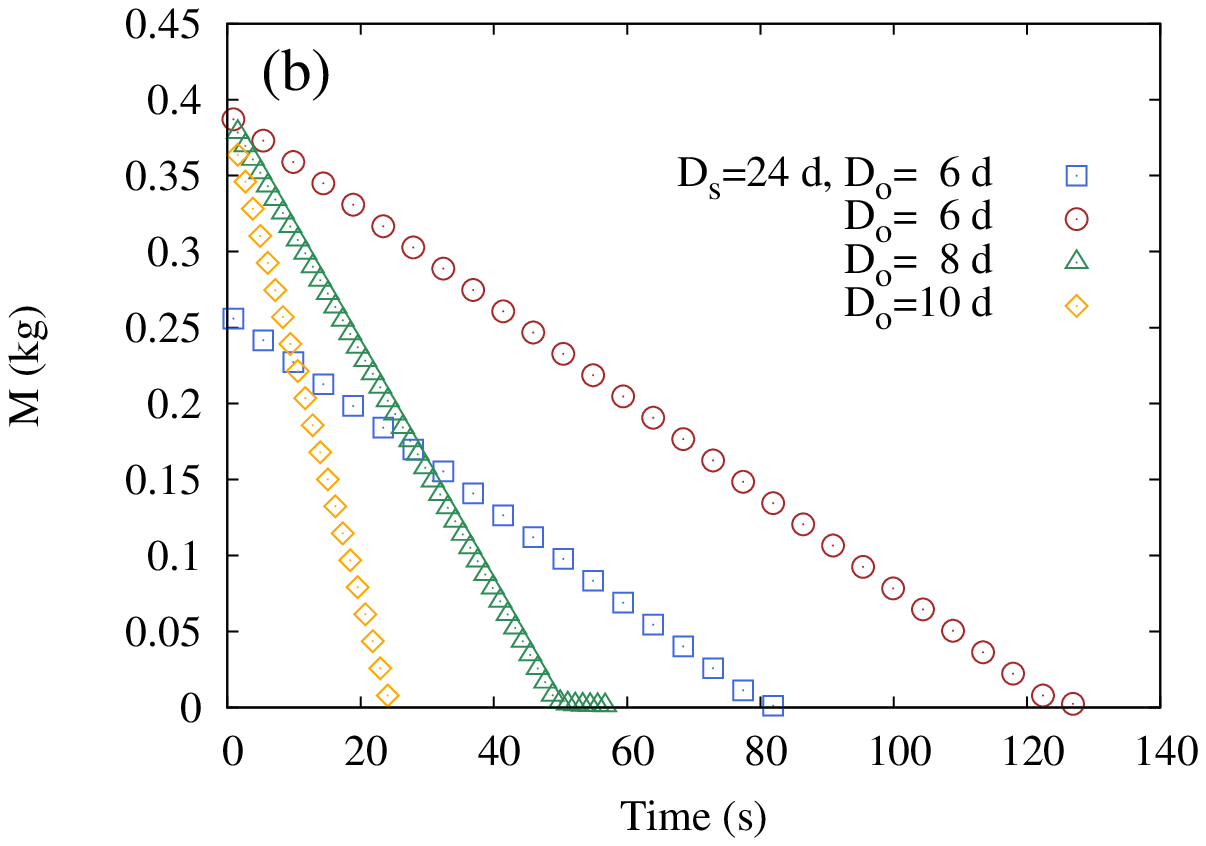}
 \end{center}
\caption{(a) Mass flow rate as a function of the total mass $M$ in the silo during the discharge. The flow rate is scaled by the value from the Beverloo equation (i.e., the asymptotic flow rate for a tall column). The solid line corresponds to the prediction by Eq. (\ref{diff}) using $\muI=0.2$ and $\alpha=2.5$. Symbols correspond to DEM data for different $\Do$ and $\Ds$. Unless otherwise stated $\Ds=30d$. The friction coefficient is set to $\mu=0.5$ for the particle--particle and the particle--wall interactions. {(b) Mass in the silo as a function of time for the same simulations as in part (a). Note that for the narrow silo we used a lower initial mass. }}
\label{fig-caudal-t}
\end{figure}

Equation (\ref{diff}) closed by Eqs. (\ref{eq-szzo}) and (\ref{eq-walters2}) do not need to be solved numerically as a function of time since a parametric plot of $-\dot{M}(t)$ as a function of $M(t)$  can be directly obtained. In Fig. \ref{fig-caudal-t}(a) we plot the flow rate during discharge. {We also include $M(t)$ without scaling in Fig. \ref{fig-caudal-t}(b) as a reference.} As we can see, for most part of the discharge, Eq. (\ref{diff})  predicts a constant flow rate, in agreement with our DEM simulations and experimental observations. However, the model predicts an early drop in the flow rate well before this is observed in the simulations. We believe this discrepancy is connected with the poor prediction of the pressure contribution $\msr$ for the final stages of the discharge (see Fig. \ref{fig-presion}(a)). We expect that new developments on the estimation for the internal pressure in silos will lead to an immediate improvement of the prediction of Eq. (\ref{diff}) for the evolution of the flow rate. 

Despite the shortenings of the prediction in the final stages of the discharge, Eq. (\ref{diff}) provides, to our knowledge, the first mean to calculate the evolution of the discharge. This opens new possibilities to study problems such as forced discharges that show an non-constant flow rate \cite{madrid2018}. 

\section{Predictions for low particle--particle friction}

In the previous sections, we have observed an excellent agreement between the Beverloo equation and our theory by using $\muI=0.2$. This is possible because $\muI$ is not very sensitive to the material properties of the grains if $\mu> 0.4$ for the particle--particle interaction \cite{dacruz2005}. This is why the a wide range of materials can be fitted to the Beverloo equation with a single value for the constant $C$. However, if $\mu< 0.4$ the effective friction $\muI$ starts to drop. Therefore, we must expect that the flow rate will depend on the particle--particle friction if  $\mu< 0.4$. Some authors have indeed reported that for low friction the flow rate is higher \cite{kondic2014}. This effect is not accounted for in the Beverloo equation except for the fact that the constant $C$ can be fitted to a new value. However, Eq. (\ref{diff}) can predict the effect of lowering $\mu$ without changing neither the non-dimensional prefactor nor $k$ (we recall that $k$ is introduced when correcting $\Do$ by $\Do-kd$). The correction is obtained by tunning $\muI$, which has a clear physical interpretation from the $\muI$-rheology.     

\begin{figure}
 \begin{center}
  \includegraphics[width=0.8\columnwidth]{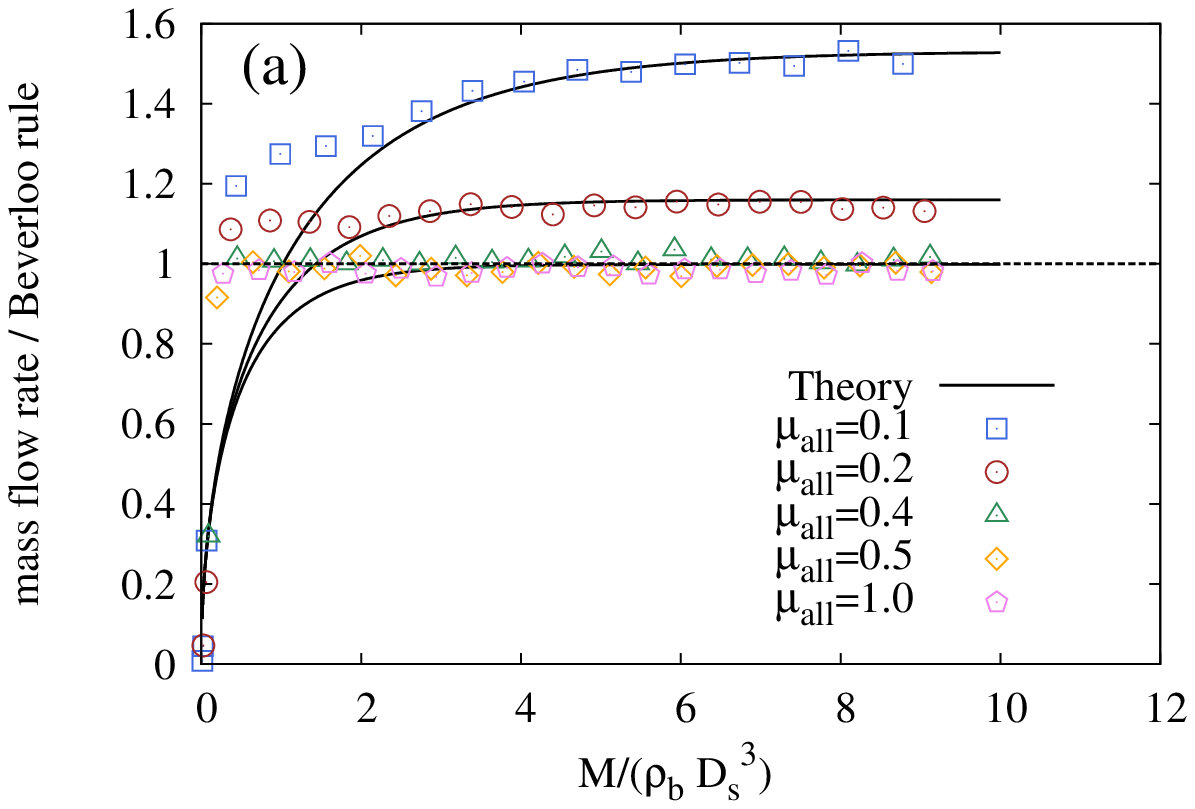}\\
  \includegraphics[width=0.8\columnwidth]{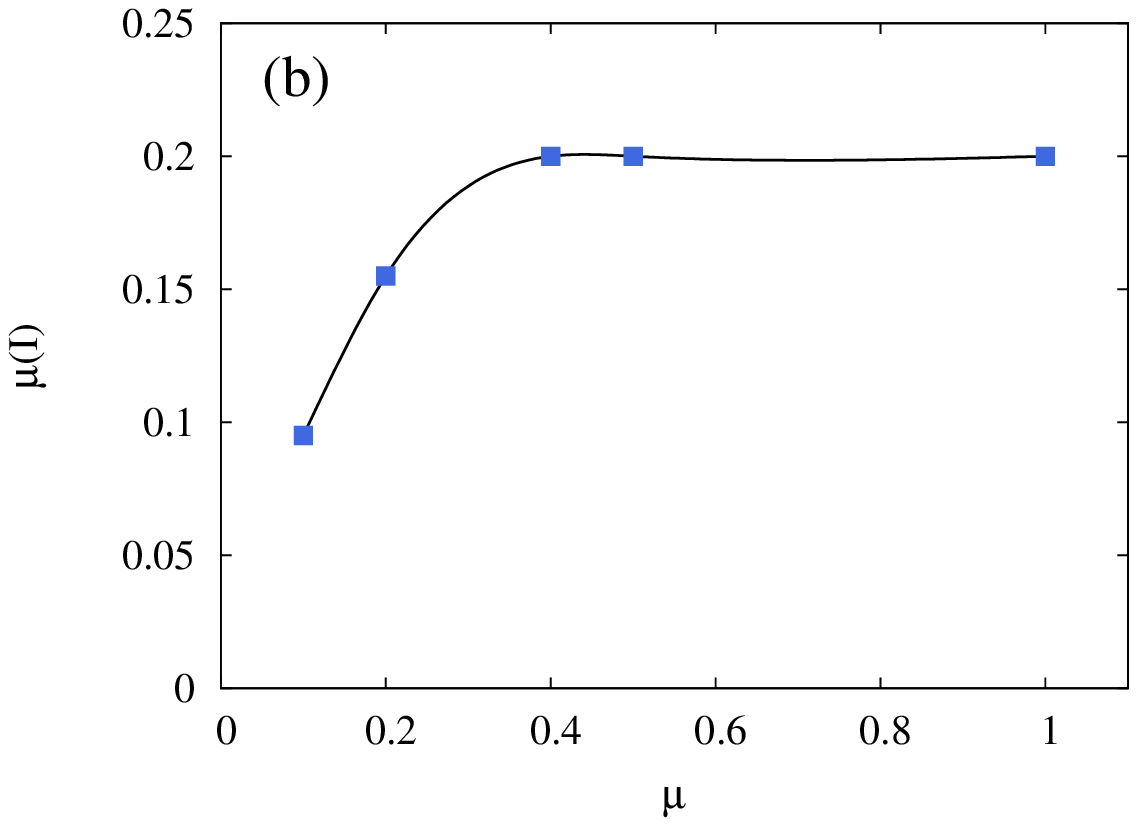}\\
    \includegraphics[width=0.8\columnwidth]{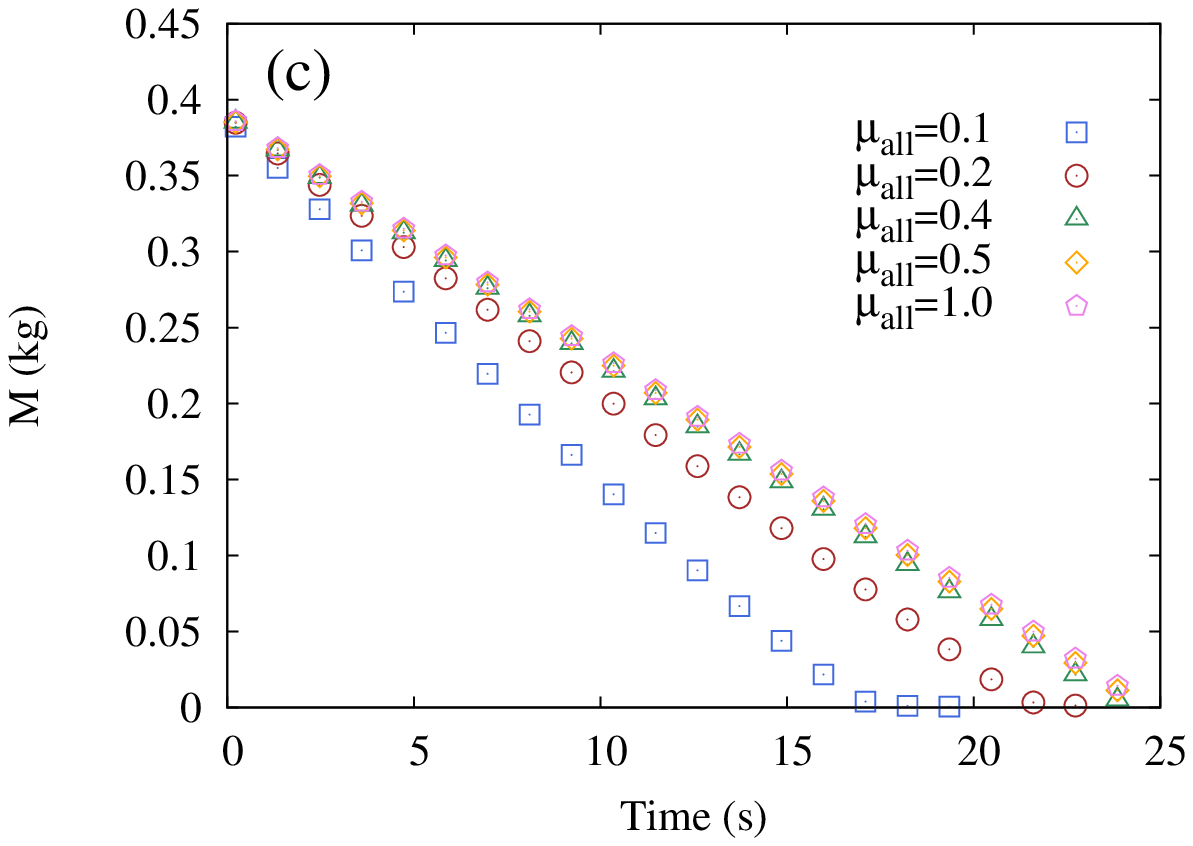}
 \end{center}
\caption{(a) Mass flow rate as a function of mass in the silo for DEM simulations for different friction coefficients $\mu$. We set $\Do=10 d$ and $\Ds=30d$. The particle--particle and particle--wall values of $\mu$ are made equal. The flow rate is scaled by the Beverloo equation. The solid lines correspond to the prediction of Eq. (\ref{diff}) where $\muI$ is set to $0.2$ if $\mu>0.4$ and to lower values if $\mu<0.4$. (b) Values of $\muI$ as a function of $\mu$ used in part (a). The solid line in part (b) is only a guide to the eye. {(c) Mass in the silo as a function of time for the same simulations as in part (a).}}
\label{fig-mu}
\end{figure}

In Fig. \ref{fig-mu}(a), we plot the mass flow rate during discharge for different values of $\mu$ used in our DEM simulations. We scaled the flow rate by the one predicted by Beverloo in the previous sections, which holds valid for large friction ($\mu>0.5$). The solid lines correspond to Eq. (\ref{diff}), where we have used as a fitting parameter the value of $\muI$. For $\mu> 0.4$ we use $\muI=0.2$ as in the previous sections. For $\mu< 0.4$ we set $\muI$ to lower values, while we keep $\alpha^{-1}=\tan\phi$ and $k=1.72$ as in the previous section. The actual values of $\muI$ used are shown in Fig. \ref{fig-mu}(b). {For reference, we also include $M(t)$ without scaling in Fig. \ref{fig-mu}(c).}

The values of $\muI$ are difficult to predict. These depend, for example, on the geometry. In the quasistatic limit for plane shear, while $\muI=0.28$ in 2D \cite{dacruz2005} it rises to $0.36$ in 3D \cite{azema2014}. It is difficult to provide an independent estimate for $\muI$ as a function of $\mu$ for the particular case of a cylindrical silo. However, Eq. (\ref{diff}) provides with the new insight that the mass flow rate scales as $\sqrt{\muI}$.

\section{Conclusions}

We have used the work--energy theorem to derive an expression for the mass flow rate of a discharging silo as a function of the mass inside the silo [see Eq. (\ref{diff})]. For wide silos and stiff grains, we have shown that the changes in kinetic energy and elastic energy can be neglected. We have used the concepts of the $\muI$-rheology to calculate the power dissipated during discharge. The asymptotic limit of Eq. (\ref{diff}) resembles the well known Beverloo rule. Interestingly, the non-dimensional prefactor that we predict is within $4\%$ of the experimental fitted values. Besides, we have shown that Eq. (\ref{diff}) provides a mean to explain the higher flow rates observed in low-friction materials. 

It would be important to test the limitations of this approach when different conditions are used such as: two-dimensional silos; silos, orifices and particles with different shapes; hoppers; bumpy walls; use of an overweight; use of soft and/or deformable grains; etc. In particular, recent simulations with spherocylinders \cite{hidalgo2018} have shown that $\muI$ depends on aspect ratio, which should have an impact on the flow rate. It is also interesting the potential application to suspensions and submerged grains passing through constrictions since there are recent developments that indicate that the $\muI$-rheology is suitable to describe the flow in these systems \cite{boyer2011,Houssais2016}. Forced silo discharges using overweights have shown non-constant flow rates during discharge \cite{madrid2018,madrid2019}. An extension of Eq. (\ref{diff}) to forced flows may be suitable to model the discharge under such conditions.     

Finally, we suggest a potential extension for this theory. The $\muI$-rheology is a local approximation that does not account for the effects of particle size. The use of non-local approximations (see for example \cite{kamrin2012}) may help to obtain a correction to account for the ``empty annulus'' effect and so avoid the simplistic approximation introduced to consider the boundary effects at the orifice.

\appendix
\section{Interactions for the DEM simulations}

We used LIGGGHTS \cite{liggghts} to calculate the trajectories 
of each particle by integrating the Newton-Euler equations \cite{poschel2005}. The equations of motion for $N$ grains are solved via a velocity-Verlet algorithm by advancing in small time intervals $\Delta t$. We used the Hertz particle--particle interaction model with Coulomb criterion \cite{poschel2005}. In this model, the normal component of the contact force between two grains $i$ and $j$ is defined as
\begin{equation} \label{eq-fn}
 F_{\rm n}=k_{\rm n} \delta_{\rm n}^{3/2}-\gamma_{\rm n} \delta_{\rm n}^{1/4} \dot{\delta}_{\rm n},
\end{equation}
where
\begin{equation}
 k_{\rm n}=\frac{\sqrt{d}}{3}\frac{Y}{(1-\nu^2)}
\end{equation}
is the elastic constant for normal contacts and $\delta_{\rm n}$ is the overlap in the normal direction between particles $i$ and $j$. $Y$ is the Young's modulus, $\nu$ the Poisson ratio and $d$ the diameter of the particles.
The damping constant for normal contact is
\begin{equation}
 \gamma_{\rm n} =\sqrt{\frac{5}{6}} \frac{\ln(e)}{\sqrt{\ln^2(e)+\pi^2}} \sqrt{\frac{Y m}{1-\nu^2}}\sqrt[4]{d},
\end{equation}
which results from the solution of the Herztian spring-dashpot model \cite{antypov2011}. Here, $m$ is the grain mass and $e$ is de coefficient of restitution wich is independent of the velocity \cite{tsuji1993}. $\dot{\delta}_{\rm n}$ is the normal relative velocity of the particles $i$ and $j$.

The tangential component of the contact force between particles $i$ and $j$ is
\begin{equation} \label{eq-ft}
 F_{\rm t}= -\textrm{sign}(v_{\rm t}) \textrm{min}( | k_{\rm t} \delta_{\rm t} \delta_{\rm n}^{1/2}-\gamma_{\rm t} \dot{\delta}_{\rm t} \delta_{\rm n}^{1/4} | ,\mu F_{\rm n}),
\end{equation}
being $v_{\rm t}$ the tangential relative velocity of the spheres at the point of contact which takes into accounts the relative velocity of the centers of the spheres and their rotation. 
The elastic constant for tangential contacts is
\begin{equation}
 k_{\rm t}=\frac{Y \sqrt{d}}{(2-\nu)(1+\nu)}.
\end{equation}
The damping constant for tangential contact is
\begin{equation}
 \gamma_{\rm t} =-2 \sqrt{\frac{5}{6}} \frac{\ln(e)}{\sqrt{\ln^2(e)+\pi^2}} \sqrt{\frac{Y m}{2(2-\nu)(1+\nu)}}\sqrt[4]{d}.
\end{equation}
$F_{\rm t}$ is limited by Coulomb friction, being $\mu$ the friction coefficient \cite{poschel2005}. In this model, the static and dynamic friction coefficients coincide. The tangential displacement $\delta_{\rm t}$, which depends on the history of the contact, is calculated as
\begin{equation}
 \delta_{\rm t}(t)=\int^{t}_{t_{\rm c}} v_{\rm t} (t') dt',
\end{equation}
where $t_{\rm c}$ is the time at which the contact begins.

{To calculate the dissipated power in DEM simulations there exist two basic approaches as described below.

\emph{Approach 1.} At each time step the work done by the non-conservative terms in the contact forces is calculated and this is saved in a cumulative variable. This is done by taking the dot product between the displacement $\mathbf{\delta r}$ of each contact and the non-conservative part of the contact force in that time step [i.e., the force vector resulting from combining the second term in Eq. (\ref{eq-fn}) and either, the second term in the absolute value of Eq. (\ref{eq-ft}) or $\mu F_{\rm n}$, whichever applies]. This approach has the advantage that one can accumulate separately different contributions such as the tangential contribution [frictional, Eq. (\ref{eq-ft})] and the normal contribution [inelastic collision, Eq. (\ref{eq-fn})]. However, this is CPU demanding since the operation has to be done at every time step of the simulation.

\emph{Approach 2.} At two different arbitrary times (which may be separated by many time steps) one calculates the potential energy of all conservatives forces and the total kinetic energy (including rotations). The conservative forces include gravity and the conservative terms of the contact forces [i.e., the first term in Eq. (\ref{eq-fn}) and the first term in the absolute value of Eq. (\ref{eq-ft})]. These potential energies can be calculated at any time since the conservative forces depend only on the current positions of the particles (not the actual trajectories nor velocities). Therefore, one does not need to make this calculation at every time step to track trajectories. The difference of total energy (potential + kinetic) between the two times under consideration corresponds to the energy dissipated by the non-conservative terms in the contact interactions. Unfortunately, this approach will not provide detailed information on the contribution of each dissipation mode (friction and normal collisions).

We have used \emph{Approach 2} to obtain the dissipated energy. This prevents us from having access to the detail of how much energy is lost by friction and how much by normal collisions. However, this is a very efficient method and suffices for the purposes of the current study.}

\section{Contributions to the work--energy theorem}

In Fig. \ref{energies}, we plot the different contributions to the power injected or extracted from the silo obtained via a DEM simulation of the silo discharge. We have run simulations for various silos diameters, particle--particle interaction parameters and orifice sizes. All cases studied display the same trends as the ones shown in this sample simulation.

\subsection{Internal kinetic energy ($\dot{K}_\textrm{in}$)}

\begin{figure}
 \begin{center}
  \includegraphics[width=0.9\columnwidth]{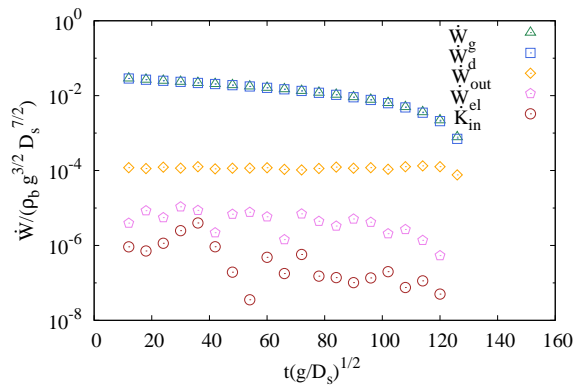}
 \end{center}
\caption{Contributions to the work--energy balance during the discharge of a silo as a function of time (DEM simulations). Power injected by gravity $\dot{W}_{\rm g}$, {dissipated power $\dot{W}_{\rm d}$,} power loss through the orifice $\dot{W}_{\rm out}$, elastic power $\dot{W}_{\rm el}$ and rate of change of the internal kinetic energy $\dot{K}_{\rm in}$. {Data corresponds to a silo with $\Ds=30d$ and $\Do=6d$ while the particle--particle and particle--wall friction coefficients are set to $\mu=0.5$.}}
\label{energies}
\end{figure}

As we can see in Fig. \ref{energies}, the contribution of $\dot{K}_\textrm{in}$ is one order of magnitude smaller than $\dot{W}_{\rm out}$ and three orders of magnitude smaller than $\dot{W}_{\rm g}$. Therefore, we can neglect $\dot{K}_\textrm{in}$. To provide a partial explanation for this observation consider the kinetic energy of the grains inside the silo \cite{madrid2017b}
\begin{equation}
 K_\textrm{in}(t)=\frac{1}{2}\sum_{i=1} m \bm{v}_i^2(t).
\end{equation}
The sum runs over all particles inside the silo at time $t$ and $\bm{v}_i(t)$ is the velocity of particle $i$. This can be expressed in terms of the center of mass velocity $v_\textrm{cm}$  and the ``granular temperature'' as $K_\textrm{in}(t)= K_\textrm{in}^\textrm{cm}(t) + K_\textrm{in}^\textrm{temp}(t)$, being
\begin{equation} \label{kinetic-cm2}
 K_\textrm{in}^\textrm{cm}(t)=\frac{1}{2}M(t) v_\textrm{cm}^2(t),
\end{equation}
and
\begin{equation} \label{kinetic-cm3}
 K_\textrm{in}^\textrm{temp}(t)= \frac{1}{2}\sum_{i=1}^{N(t)}m_i [\bm{v}_i(t)-\bm{v}_\textrm{cm}(t)]^2.
\end{equation}

We will disregard the ``temperature'' term and focus on the center of mass. We assume here that the $x$ and $y$ components of $\bm{v}_\textrm{cm}$ are null and therefore  $|\bm{v}_\textrm{cm}|=v_\textrm{cm}=-\dot{z}_\textrm{cm}$. Hence, the rate of change of the kinetic energy is

\begin{align}
\dot{K}_\textrm{in}(t)&\approx \dot{K}_\textrm{in}^\textrm{cm}(t) \notag\\&=M(t) v_\textrm{cm}(t) \dot{v}_\textrm{cm}(t) + \frac{1}{2}v_\textrm{cm}^2(t) \dot{M}(t), 
\end{align}
which can be written, using Eqs. (\ref{equiv}) and (\ref{q-vcm}), as
\begin{equation}
 \dot{K}_\textrm{in}(t)\approx \frac{1}{4\rho^2_\textrm{b} A^2_\textrm{s}}
 \left[M(t)\dot{M}(t)\ddot{M}(t)+\frac{1}{2}\dot{M}^3(t)\right] \label{k-in}.
\end{equation}

As we can see, $\dot{K}_\textrm{in}$ decays as $A_\textrm{s}^{-2}$. Therefore, $\dot{K}_\textrm{in}$ will be small for wide silos. We will see below that other terms in the energy balance decrease as $A_\textrm{s}^{-1}$.

\subsection{Gravitational energy ($\dot{W}_\textrm{g}$)}

The gravitational potential energy of the particles inside the silo is
\begin{equation}
 U_\textrm{g}(t)=M(t)gz_\textrm{cm}(t)=\frac{g M^2(t)}{2\rho_\textrm{b} A_\textrm{s}}.
\end{equation}
Where we have used Eq. (\ref{equiv}). Therefore, the power injected by the action of gravity is

\begin{equation}
 \dot{W}_\textrm{g}(t)=-\dot{U}_\textrm{g}(t)=-\frac{g}{\rho_\textrm{b} A_\textrm{s}} M(t) \dot{M}(t). \label{wg}
\end{equation}

As we can see, $\dot{W}_\textrm{g}$ scales with $A_\textrm{s}^{-1}$ in contrast to the faster decay displayed by $\dot{K}_\textrm{in}$. We recall here that $\dot{M}(t)$ is negative; therefore $\dot{W}_\textrm{g}(t)$ is positive. {It is worth mentioning that $\dot{W}_\textrm{g}$ is mostly dissipated and goes into $\dot{W}_{\rm d}$ (see Fig. \ref{energies}). A very small portion of $\dot{W}_\textrm{g}$ goes into the draining grains ($\dot{W}_\textrm{out}$), as we discuss in the following subsection.}

\subsection{Discharge energy loss ($\dot{W}_\textrm{out}$)}

While the system discharges, the particles that leave the system take away some energy since their own kinetic energy is no longer part of the internal energy of the silo. If during a time interval $dt$ at time $t$ the system discharges a mass $dM= -\dot{M}(t)dt$ at velocity $v_{\rm out}$, then the kinetic energy removed per unit time is

\begin{equation}
\dot{K}_\textrm{out}(t) = -\frac{1}{2} v_\textrm{out}^2 \dot{M}(t).
\end{equation}

By definition, $-\dot{M}(t)= \rho_\textrm{o} A_\textrm{o} v_\textrm{out}$. Therefore,
\begin{equation}
 \dot{W}_\textrm{out}=\dot{K}_\textrm{out}(t) =-\frac{\dot{M}^3(t)}{2\rho^2_\textrm{o} A^2_\textrm{o}}. \label{wout}
\end{equation}

The power lost by discharge does not depend on the silo cross section but on the orifice cross section $A_\textrm{o}$. Then again, $\dot{W}_\textrm{out}$ is positive since $\dot{M}(t)$ is negative. 

{It is important to note that the definition $-\dot{M}(t)= \rho_\textrm{o} A_\textrm{o} v_\textrm{out}$ is not strictly correct since the velocity and the density at the orifice are not homogeneous (see Appendix C for details) \cite{janda2012}.}

\subsection{Elastic energy ($\dot{W}_\textrm{el}$)}

If the grains are stiff, the variation in the elastic energy at the contacts are expected to be small. $\dot{W}_\textrm{el}$ corresponds to the rate of change of the conservative component of the contact forces $\dot{E}_\textrm{el}$. In Fig. \ref{energies}, we show the rate of change in elastic energy during a silo discharge in a DEM simulation. As we can see, this term is of the order of the change in kinetic energy $\dot{K}_{\rm in}$ and can be neglected in comparison with $\dot{W}_{\rm g}$ and $\dot{W}_{\rm out}$. Of course, this may be inadequate for very soft grains. For our purposes, we simply disregard this contribution in the analysis. However, this can be eventually included by using the stress based expression for the elastic energy density in the limit of small deformations \cite{landau}  

\begin{equation}
 E_\textrm{el} = A_\textbf{s} z_\text{cm} \bm{\sigma} \bm{\epsilon}, \label{elastic}
\end{equation}
being $\bm{\sigma}$ the stress tensor and $\bm{\epsilon}$ the strain tensor.

{\section{Estimation of $\rho_{\rm o}$}

Since the packing fraction and the particle velocities are not homogeneous across the orifice, the correct flow rate has to be calculated as \cite{janda2012}

\begin{equation}\label{eq-Q-def}
Q = \int_{0}^{\Do/2} {v_{\rm o}(r) \rhoo(r) 2 \pi r dr},
\end{equation}
where $v_{\rm o}(r)$, and $\rhoo(r)$ are the vertical velocity and density profiles at the horizontal plane of the orifice as a function of the distance $r$ to the center of the orifice.

If the profiles where flat, one can write 

\begin{equation}\label{eq-Q-simple}
Q = \rhoo v_{\rm o} \Ao, 
\end{equation}
where $\Ao$ is the cross section of the orifice and $\rhoo$ and $v_{\rm o}$ are the mean density and mean vertical velocity at the orifice, i.e., 

\begin{align}\label{eq-mean-v-and-rho}
\rho_{\rm o} = \Ao^{-1} \int_{0}^{\Do/2} {\rhoo(r) 2 \pi r dr},\notag\\
v_{\rm o} = \Ao^{-1} \int_{0}^{\Do/2} {v_{\rm o}(r) 2 \pi r dr}.
\end{align}

This is in general invalid (for non-flat profiles) because the integral of the product $v_{\rm o}(r)\rhoo(r)$ is different from the product of the integrals of $v_{\rm o}(r)$ and $\rhoo(r)$. It is worth mentioning that if $\rhoo(r)$ does not depend on $r$, then Eq. (\ref{eq-Q-simple}) is valid if $v_{\rm o}$ is defined as in Eq. (\ref{eq-mean-v-and-rho}). This is why expression (\ref{eq-Q-simple}) is valid for incompressible fluids. However, for granular materials, the density across the orifice is not constant and Eq. (\ref{eq-Q-simple}) is actually a poor approximation if $\rhoo$ and $v_{\rm o}$ are defined as in (\ref{eq-mean-v-and-rho}). 

Therefore, the definitions for $v_{\rm o}$ and $\rhoo$ in the simplified Eq. (\ref{eq-Q-simple}) need to be replaced by ``effective values'' rather than ``mean values''. As an example, we can set $v_{\rm o}$ to the mean value of $v_{\rm o}(r)$ but then chose $\rhoo$ to yield the correct flow rate. We have done this for experiments and simulations from the literature and found that setting $v_{\rm o}$ to the mean value of $v_{\rm o}(r)$ leads to an effective value of $\rhoo \sim \rhob/2$. The mean value for the packing fraction at the orifice is in fact lower (about $\rhob/3$). As an example, in our simulations for $\Do=6d$, the mass flow rate is $Q=0.00312$ kg/s. The corresponding vertical velocity at the orifice (averaged over a cylindrical region that fits exactly in the orifice and has a height of $d$) is $v_{\rm o}=0.1353$ m/s. Therefore, from Eq. (\ref{eq-Q-simple}) above, we find that the effective apparent density at the orifice is $\rhoo=815$ kg/m$^3$. Since our bulk $\rhob=1475$ kg/m$^3$, then $\rhoo/\rhob = 0.55$. 
}

\begin{acknowledgments}
This work has been supported by ANPCyT (Argentina) through grant PICT-2012-2155, Universidad Tecnol\'ogica Nacional (Argentina) through grant  PID-MA0FALP0002184 and PID IFI1871, Centro Argentino Franc\'es de Ciencias de la Ingenier\'ia (CAFCI, Argentina-Francia) and FONACIT through grant 2015000072 (INVUNI2013-1563) Universidad Sim\'on Bol\'ivar (Venezuela). 
\end{acknowledgments}


\begin{thebibliography}{50}

\bibitem{hagen1852} G. H. L. Hagen, Bericht \:uber die zur Bekanntmachung geeigneten Verhandlungen der K\:oniglich Preussischen Akademie der Wissenschaften zu Berlin. 35--42 (1852).

\bibitem{tighe2007} B. P. Tighe and M. Sperl, \textit{Gran. Matt.} \textbf{9}, 141 (2007)

\bibitem{BrownBook} R. L. Brown and J. C. Richards, \textit{Principles of Powder Mechanics}, Pergamon Press, Oxford, (1970).

\bibitem{beverloo} W. Beverloo, H. Leniger and J. Van de Velde, Chem. Eng. Sci. {\bf 15}, 260 (1961).

\bibitem{NeddermanBook} R. M. Nedderman, {\it Statics and kinematics of granular materials} (Cambridge University Press, Cambridge, 2005).


\bibitem{madrid2018} M. A. Madrid, J. R. Darias and L. A. Pugnaloni, {\it Forced flow of granular media: Breakdown of the Beverloo scaling}, Europhys. Lett. {\bf 123}, 14004 (2018).

\bibitem{nedderman1982} R. M. Nedderman, U. T\"{u}z\"{u}n, S.B. Savage and G.T. Houlsby, \textit{J. Chem. Eng. Sci.} \textbf{37}, 1597-1609 (1982)

\bibitem{tuzun} U. T\"{u}z\"{u}n, G. T. Houlsby, R. M. Nedderman and S. B. Savage, \textit{J. Chem. Eng. Sci.}\textbf{37}, 1691-1709 (1982)

\bibitem{savage} S. B. Savage, R. M. Nedderman, U. T\"{u}z\"{u}n and G.T. Houlsby, \textit{J. Chem. Eng. Sci.} \textbf{38}, 189-195 (1983).

\bibitem{ristow} G. H. Ristow, \textit{Physica A} \textbf{235}, 319 (1997).

\bibitem{mankoc2007} C~Mankoc, A~Janda, R~Ar\'evalo, J~Pastor, I~Zuriguel, A~Garcimart\'in and  D~Maza, Granular Matter {\bf 9}, 407 (2007).
 
\bibitem{kondic2014} L Kondic, {\it Simulations of two dimensional hopper flow}, Granular Matter {\bf 16}, 235   (2014).

\bibitem{wilson2014}  T J Wilson, et al., {\it Granular discharge rate for submerged hoppers}, Papers in Physics {\bf 6}, 060009 (2014).

\bibitem{goldberg2016} E. Goldberg, C. M. Carlevaro and L. A. Pugnaloni, Papers in Physics \textbf{7}, 070016 (2015).

\bibitem{villagran} M. C. Villagr\'an Olivares, J. G. Benito, R. O. U\~nac and A. M. Vidales, {\it Towards a one parameter equation for a silo discharging model with inclined outlets}, Powder Technology \textbf{336} 265 (2018).

\bibitem{rubio2015} S. M. Rubio-Largo, A. Janda, D. Maza, I. Zuriguel and R. C. Hidalgo, Phys. Rev. Lett.  \textbf{114}, 238002 (2015).

\bibitem{mcdougall1965} I. R. McDougall and A. C. Evans, Rheologica Acta \textbf{4}, 218 (1965).


\bibitem{staron2012} L. Staron, P.-Y- Lagr\'ee \ S. Popinet, Phys. Fluids {\bf 24}, 103301 (2012).

\bibitem{dacruz2005} F. da Cruz, S. Emam, M. Prochnow, J. N. Roux and F. Chevoir, Phys. Rev. E {\bf 72}, 021309 (2005).


\bibitem{liggghts}  C. Kloss, C. Goniva, A. Hager, S. Amberger and S. Pirker, \textit{Prog. Comput. Fluid Dynamics}, \textbf{12}, 140 (2012).

\bibitem{azema2014} E. Az\'ema and F. Radjai, Phys. Rev. Lett. {\bf 112}, 078001 (2014).

\bibitem{foot} Note that we have selected the negative root since this is the physically meaningful root corresponding to negative $\dot{M}$.

\bibitem{walters1973} J. K. Walters, A theoretical analysis of stresses in silos with vertical walls, Chem. Eng. Sci. {\bf  28}, 13 (1973).

\bibitem{walker1966} D. M. Walker, An approximate theory for pressures and arching in hoppers, Chem. Eng. Sci. {\bf 21}, 975 (1966).


\bibitem{fowler1959} {R. T. Fowler and J. R. Glastonrury, Chem. Eng. Sci. \textbf{10}, 150 (1959).} 

\bibitem{janda2012} A. Janda, I. Zuriguel and D. Maza, Phys. Rev. Lett. {\bf 108}, 248001 (2012).

\bibitem{hidalgo2018} R. C. Hidalgo, B. Szab\'o, K. Gillemot, T. B\"orzs\"onyi, T. Weinhart, Phys. Rev. F {\bf 3}, 074301 (2018).

\bibitem{boyer2011} F. Boyer, E. Guazzelli and O. Pouliquen, Phys. Rev. Lett. {\bf 107}, 188301 (2011).

\bibitem{Houssais2016} M. Houssais, C. P. Ortiz, D. Durian and J. Jerolmack, Phys. Rev. E {\bf 94}, 062609 (2016).


\bibitem{madrid2019} M. A. Madrid and L. A. Pugnaloni, Granular Matt. {\bf 21}, 76 (2019).

\bibitem{kamrin2012} K. Kamrin and G. Koval, {\it Nonlocal constitutive relation for steady granular flow}, Phys. Rev. Lett. {\bf 108}, 178301 (2012).

\bibitem{poschel2005} T. P\"oschel and T. Schwager,  {\it Computational Granular Dynamics. Models and Algorithms}, Springer, Berlin (2005).

\bibitem{antypov2011} D. Antypov and J. Elliott, Europhys. Lett.  \textbf{94}, 50004 (2011).

\bibitem{tsuji1993} Y. Tsuji and T. Kawaguchi, T. Tanaka, Powder Technol. \textbf{77}, 79 (1993). 


\bibitem{madrid2017b} M. A. Madrid, J. R. Darias and L. A. Pugnaloni, EPJ Web of Conf. {\bf 140}, 03041 (2017).

\bibitem{landau} L. D. Landau and E. M. Lifshitz, {\it Theory of Elasticity} (3rd ed.). Butterworth Heinemann, Oxford (1986).


\end{thebibliography}
\end{document}